# Learning Cybersecurity vs. Ethical Hacking: A Comparative Pathway for Aspiring Students

Author: Fahed Quttainah

## Abstract

**Learning Cybersecurity vs. Ethical Hacking: A Comparative Pathway for Aspiring Students** provides a comprehensive overview of two distinct yet interrelated fields in the realm of digital security. Geared toward beginners and students considering a career in this area, the paper defines **cybersecurity** and **ethical hacking**, clarifying how they differ in scope, purpose, and methodology. Cybersecurity is presented as a broad discipline focused on defending systems and data, whereas ethical hacking is a specialized practice of offensively testing those systems to uncover vulnerabilities. The discussion compares the educational pathways and training approaches for each field, from formal academic programs to hands-on certifications, and outlines the key technical skills and professional certifications that employers value in cybersecurity and ethical hacking roles. Common career paths are described - ranging from security analysts and engineers to penetration testers and red team specialists - along with the industries that employ these professionals. Finally, the paper offers guidance to help aspiring students choose the path best suited to their interests and goals. Through this comparative analysis, readers will gain a clear understanding of what each pathway entails and how to navigate their educational and professional development in cybersecurity or ethical hacking.

## Introduction

In today's digital age, protecting sensitive information and securing computer systems has become a critical priority for organizations across all sectors. As cyber threats grow in sophistication and frequency, the demand for skilled professionals who can safeguard digital assets is at an all-time high. Two popular career pathways have emerged in response to this need: **cybersecurity** and **ethical hacking**. While both paths ultimately contribute to the security of information systems, they do so in different ways and require distinct mindsets. However, for students and newcomers to the field, the distinction between cybersecurity and ethical hacking can be unclear. Are they the same, or is ethical hacking just a part of cybersecurity? What does each role involve day-to-day, and what knowledge or training is needed to excel in one versus the other? These are important questions for anyone considering a future in the field of information security.

This paper seeks to clarify the differences and overlaps between cybersecurity and ethical hacking as career paths. It begins by defining the core concepts of each field and explaining their objectives and methodologies. Next, it compares the academic routes and training opportunities available for learning cybersecurity versus those for learning ethical hacking. The discussion then highlights common job roles associated with each path and the types of industries where these skills are applied. An examination of the skill sets and professional certifications relevant to cybersecurity and ethical hacking follows, illustrating what competencies aspiring professionals are expected to develop. Finally, the paper offers advice on how students can evaluate their personal interests and strengths to choose the pathway that best aligns with their goals. By providing a side-by-side comparison of these two pathways, the aim is to equip readers with the knowledge to make an informed decision about starting a career in defending cyberspace, whether as a broad-based cybersecurity expert or as a specialized ethical hacker.

## Defining Cybersecurity

**Cybersecurity** is the broad practice of protecting computers, networks, software, and data from unauthorized access, theft, damage, or disruption. In essence, cybersecurity encompasses all the strategies and measures used to prevent or mitigate cyber attacks and to ensure the confidentiality, integrity, and availability of information. This field covers a wide range of defensive techniques and domains, including network security, application security, information governance, cryptography, cloud security, and more. A cybersecurity professional's primary mission is to **defend**: they design and implement safeguards like firewalls, intrusion detection systems, access controls, and encryption protocols to create a secure environment. They also establish policies and incident response plans so that if a breach or threat is detected, it can be contained and resolved with minimal damage. Cybersecurity experts work proactively to anticipate potential threats (such as malware outbreaks, phishing campaigns, or insider misuse) and to strengthen systems against those threats before incidents occur.

In practical terms, cybersecurity roles involve monitoring for suspicious activities, analyzing system vulnerabilities, and continuously updating defenses. For example, a security analyst might review alerts from a monitoring system to identify possible intrusions, whereas a security engineer might develop a new secure network architecture for a company. **Risk management** is another core aspect of cybersecurity - professionals assess the likelihood and impact of various security risks and then prioritize security improvements accordingly. Importantly, cybersecurity is not purely a technical arena; it also involves **compliance** and **policy**. Organizations must adhere to regulations and standards (like GDPR for data protection or industry-specific security standards), so cybersecurity professionals often ensure that security practices meet these legal and ethical requirements. In summary, cybersecurity is a comprehensive, defensive field dedicated to guarding digital assets and maintaining trust in information systems.

## Defining Ethical Hacking

**Ethical hacking**, also known as *penetration testing* or *white-hat hacking*, is a specialized subset of cybersecurity that focuses on identifying and exploiting vulnerabilities in systems **with permission and for defensive purposes**. Ethical hackers are professionals who use the same tools and techniques as malicious hackers (black-hat hackers) but in an authorized and lawful manner, with the goal of discovering security weaknesses before real attackers do. In other words, rather than primarily building defenses, an ethical hacker's mission is to **challenge** and **test** those defenses. By simulating realistic cyberattacks on a network, application, or device, ethical hackers reveal where security measures are insufficient. They then report these findings to the organization's security team and recommend fixes to strengthen the system.

The core idea behind ethical hacking is proactive security improvement. Companies and government agencies hire ethical hackers (often calling them penetration testers or red team specialists) to conduct controlled attacks on their infrastructure. These experts might attempt to crack passwords, bypass security controls, inject malicious code, or physically test building security - all depending on the scope of the test agreed upon. Throughout the process, ethical hackers must adhere to **strict ethical guidelines**: they only target systems they have explicit permission to test, they respect privacy by not abusing the data they may access, and they disclose all discovered vulnerabilities responsibly. Because of this ethical framework, the work of ethical hackers is sometimes summarized as "legal hacking" that helps organizations harden their defenses.

While ethical hacking is part of the cybersecurity umbrella, it requires a distinct mindset. Ethical hackers adopt the perspective of an attacker, thinking creatively about how to break into systems. They often develop an adversarial way of problem-solving - for instance, finding unconventional ways to exploit a minor flaw into a major breach. This offensive approach complements the defensive stance of general cybersecurity: the vulnerabilities identified by ethical hackers inform cybersecurity professionals where to focus their protective efforts. Ultimately, ethical hacking serves the same ultimate goal as cybersecurity (protecting systems and data), but it approaches the problem from the opposite direction - by actively *finding* the holes rather than primarily *building* the walls.

## Differences in Purpose and Methodology

Despite sharing the common goal of securing digital systems, cybersecurity and ethical hacking differ in their fundamental **purpose** and **methodology**. Cybersecurity's purpose is *preventative* and *protective*: professionals in this field aim to prevent attacks from happening and to shield data from compromise. Their success is measured by how well they can thwart or mitigate threats and maintain the integrity of an organization's information infrastructure. In contrast, the purpose of ethical hacking is *diagnostic* and *revelatory*: ethical hackers seek to expose weaknesses and demonstrate how a malicious actor could exploit them. The value of ethical hacking lies in uncovering problems so they can be fixed, thereby strengthening security postures. In short, cybersecurity asks, "How do we stop attackers and keep our systems safe?" while ethical hacking asks, "Where are the weak points an attacker could find, and what can they achieve by exploiting them?"

The methodologies employed by each field reflect this difference in objectives. **Cybersecurity professionals take a defensive approach.** They concentrate on architecture and processes: installing security tools (like antivirus software, intrusion prevention systems, and secure authentication mechanisms), maintaining system patches and updates, configuring networks safely, and enforcing security policies among users. Their work often involves continuous monitoring, incident response drills, and aligning security practices with frameworks or standards (such as ISO 27001 or NIST guidelines). A cybersecurity team might, for example, develop a comprehensive security policy, conduct employee training on safe computing practices, and actively watch network traffic for signs of intrusion. When incidents occur, they investigate and coordinate the response to minimize damage and recover normal operations quickly.

On the other hand, **ethical hackers take an offensive approach** in their methodology. They operate like real attackers would: gathering information about targets, scanning for open ports or weaknesses, and attempting to exploit vulnerabilities in a controlled manner. For instance, an ethical hacker might use tools like **network scanners** to map out which services a server is running, then use an **exploitation framework** (such as Metasploit) to attempt known exploits on those services. They might also carry out **social engineering** tactics - like phishing simulations - to test an organization's human defenses. Throughout an ethical hacking engagement, meticulous notes are kept on how each vulnerability was discovered and exploited. At the conclusion, ethical hackers typically compile a detailed report for the organization, describing each vulnerability found, the methods used to exploit it, and remediation recommendations to close the security gaps.

Another key methodological difference is the perspective on system access. Cybersecurity professionals usually operate with **authorized access** to systems as administrators or security officers; their job is to use that access to set up protections and monitor logs. Ethical hackers, by contrast, often start with **no privileged access** and must figure out a way to break in (just as an external attacker would). This means ethical hackers must be deeply knowledgeable about hacking techniques and creative in their approach, whereas cybersecurity defenders must be thorough in designing robust systems and processes. Both approaches are complementary: the defender's perspective is broad and preventive, and the attacker's perspective is narrow and probing. When combined, they provide a more complete security strategy. In practice, many organizations integrate ethical hacking (through periodic penetration tests or bug bounty programs) into their overall cybersecurity strategy, ensuring that defensive measures are regularly validated and improved based on the findings.

## Educational Pathways and Training Approaches

Preparing for a career in cybersecurity or ethical hacking can involve different educational routes, although both paths benefit from a strong foundation in computer science and information technology. **Cybersecurity education** is often pursued through formal academic programs. Many universities and colleges now offer dedicated degrees in cybersecurity or information security at the bachelor's and master's level. In such programs, students learn broad principles of computer networks, operating systems, cryptography, risk management, and security policy. Courses might cover topics like secure software development, digital forensics, incident response, and security compliance (Yeung, 2025). A traditional degree pathway provides a comprehensive understanding of the field and often includes theoretical knowledge alongside practical lab exercises. Additionally, some students enter cybersecurity roles after obtaining degrees in related fields such as computer science, software engineering, or information systems, later complementing their education with specialized security certifications or graduate certificates. Formal education in cybersecurity typically emphasizes not only technical skills but also problem-solving, analytical thinking, and an understanding of the legal and ethical context of security work (Quttainah, 2025).

In contrast, **ethical hacking training** tends to be more narrowly focused and skills-based. Because ethical hacking is a specialization, aspiring ethical hackers often seek out targeted training programs such as short courses, bootcamps, or certification-focused workshops (Cyberscourse, 2025). For example, there are intensive courses specifically geared towards penetration testing techniques, where students practice exploiting various systems in controlled lab environments (often using tools like Kali Linux, which is a popular operating system preloaded with hacking tools). Some universities include ethical hacking or offensive security as part of a broader cybersecurity curriculum, but many professionals in this niche learn through self-study and practical experience. They might participate in online "capture the flag" challenges, hacking competitions, or contribute to open-source security projects to hone their skills. Importantly, ethical hacking education is highly hands-on: learners are expected to actually attempt attacks in sandboxed settings to understand how exploits work in practice.

**Certifications and professional courses** also play a significant role in both pathways (TheAI Academy, 2025). In cybersecurity, students and professionals often obtain certifications like CompTIA Security+ (an entry-level certification covering basic security concepts) or more advanced credentials such as Certified Information Systems Security Professional (CISSP) and Certified Information Security Manager (CISM), which are respected in the industry for general cybersecurity knowledge and management skills. These certifications usually require passing an exam and, in some cases, meeting an experience requirement. On the ethical hacking side, a well-known certification is the EC-Council's Certified Ethical Hacker (CEH), which provides a broad overview of hacking tools and techniques from a defender's perspective. More technically demanding certifications such as Offensive Security Certified Professional (OSCP) require candidates to perform actual penetration testing in a lab exam, demonstrating practical hacking proficiency. Many ethical hackers also pursue the GIAC series of certifications (for example, GIAC Penetration Tester or GIAC Exploit Researcher) or other vendor-neutral credentials focusing on offensive skills.

It is worth noting that the educational pathways for cybersecurity and ethical hacking often intersect. A common approach is to start with a broad education in IT or cybersecurity to build a strong base, and then specialize later. For instance, an individual might earn a degree in cybersecurity or computer science, then attend a specialized training camp to learn advanced penetration testing techniques when they decide to focus on ethical hacking. Conversely, someone might begin by learning hacking skills informally (through online courses and self-guided practice) and later pursue a formal degree or certification to gain credibility and fill knowledge gaps in areas like policy or system administration. In both fields, continuous learning is essential: the threat landscape evolves rapidly, so whether one is a cybersecurity analyst or an ethical hacker, staying current through ongoing training, workshops, and reading is part of the career. Ultimately, the choice of educational pathway may depend on the student's learning style and career aspirations - whether they prefer a structured academic environment or a more direct, hands-on training approach.

## Skills and Professional Certifications

**Technical skill sets** required for cybersecurity and ethical hacking have considerable overlap, but there are distinct emphases in each pathway. A cybersecurity professional needs a well-rounded skill set that covers the **fundamentals of IT systems** and defensive security. Important skills include a strong understanding of computer networks (knowing how data travels through the internet, how routers, switches, and firewalls work), operating system administration (especially Windows and Linux systems, since securing servers and workstations is key), and familiarity with cybersecurity tools used for defense and monitoring (such as intrusion detection systems, SIEM platforms for log analysis, antivirus and endpoint protection software, etc.). They should also be comfortable with concepts of secure system design - for example, knowledge of encryption technologies to protect data or the principles of secure software development to avoid common vulnerabilities. Analytical skills are crucial as well: cybersecurity experts often must analyze logs or traffic patterns to detect anomalies that could indicate a threat. **Risk assessment and mitigation** skills help them prioritize which security issues to address first . Additionally, soft skills like communication are vital; much of cybersecurity work involves writing reports on security posture or policies and training others in the organization on security awareness. Indeed, roles such as security consultants or managers require translating technical risks to non-technical stakeholders and ensuring that security strategies align with business goals.

Ethical hackers, while sharing many of the same foundational IT skills, place more emphasis on **deep technical knowledge of vulnerabilities and exploits**. Key skills for an ethical hacker include proficiency in one or more programming or scripting languages (such as Python, C, or JavaScript) because coding skills enable the development of custom scripts or tools to test vulnerabilities. They must understand common software vulnerabilities (like those in web applications: SQL injection, cross-site scripting, etc.) and how to find them. Mastery of specialized tools is another hallmark of the ethical hacker's skill set: for instance, using *Nmap* for network scanning, *Metasploit* for exploitation, *Burp Suite* for web security testing, or password cracking tools like *John the Ripper*. **Reverse engineering** and knowledge of operating system internals can be important for analyzing how malware works or how to exploit a particular binary. Moreover, creative thinking and problem-solving are perhaps the defining "soft" skills of a good ethical hacker - they need to think like an attacker and often must improvise novel ways to achieve an objective when straightforward methods fail. Patience and persistence are also key; testing a system thoroughly might involve trying numerous attack vectors before finding a weak point.

In terms of **professional certifications**, both cybersecurity and ethical hacking offer numerous credentials that can boost a student's credibility and job prospects. For aspiring cybersecurity professionals, entry-level certifications like **CompTIA Security+** or **Certified Cybersecurity Technician** (offered by some organizations) validate foundational security knowledge. As one advances, certifications branch into specialized or managerial domains: the **CISSP (Certified Information Systems Security Professional)** is globally recognized for covering a broad range of security topics at an experienced level, and **CISM (Certified Information Security Manager)** focuses on security strategy and management for those looking to move into leadership. There are also certifications for specific areas, such as **Certified Cloud Security Professional (CCSP)** for cloud security or various GIAC certifications (Global Information Assurance Certification) that target areas like incident handling, security monitoring, or industrial control system security. These certifications usually require rigorous study and demonstrate a commitment to the field, which can be reassuring to employers.

For ethical hackers, some of the prominent certifications include the aforementioned **CEH (Certified Ethical Hacker)**, which is often an entry point into the field of penetration testing. The CEH exam tests knowledge of hacking techniques, though critics note it is heavy on memorization. Many serious practitioners aim for more hands-on certifications like **OSCP (Offensive Security Certified Professional)**, which requires candidates to actually exploit multiple machines in a controlled environment as part of a practical exam. Achieving OSCP is seen as proof of real-world penetration testing ability. Other specialized certs for ethical hackers include **OSCE (Offensive Security Certified Expert)** for advanced exploitation skills, or GIAC's **GPEN (GIAC Penetration Tester)** and **GXPN (GIAC Exploit Researcher and Advanced Penetration Tester)**. There are also certifications in niche areas like **Certified Red Team Professional (CRTP)** or **Certified Exploit Developer**, depending on how far one wants to go into the offensive security domain. It's not necessary (or practical) to earn every certification; rather, aspiring professionals usually pick those most relevant to their desired roles.

Crucially, whether one pursues cybersecurity or ethical hacking, **hands-on experience** often matters more than certificates alone. Labs, internships, or even home projects (like building a small network to secure, or setting up a personal "hacking lab" to practice on vulnerable systems) can develop real skills that certification exams might not fully capture. Employers in cybersecurity might ask about one's experience responding to security incidents or configuring specific security technologies, while employers hiring ethical hackers might ask for demonstrations of hacking methodology or solutions to hypothetical breach scenarios. Certifications serve as milestones and credibility markers, but building a strong portfolio of skills and experiences is equally important in both fields.

## Career Paths and Industry Roles

Careers in cybersecurity and ethical hacking offer a variety of roles, each with different responsibilities and work environments. **Cybersecurity careers** tend to span a broad spectrum of positions, reflecting the many aspects of protecting an organization's digital infrastructure. Common job titles include:

- **Security Analyst (or Information Security Analyst)** - professionals in this role monitor networks and systems for suspicious activities, investigate incidents, and maintain security tools.

- **Security Engineer** - focused on building and maintaining security solutions such as firewalls, encryption systems, and network defenses.

- **Security Architect** - responsible for designing the overall security architecture of an organization's IT environment.

- **Incident Responder / Forensics Analyst** - handles security breaches, investigates root causes, and helps restore operations.

- **Governance, Risk, and Compliance (GRC) Specialist** - ensures that the organization follows security policies and regulatory standards.

- **Chief Information Security Officer (CISO)** - leads the organization's entire security strategy and teams.

Cybersecurity roles are found in every sector: finance, healthcare, government, education, defense, technology, and more. Practically every organization that uses digital data needs cybersecurity professionals.

**Ethical hacking careers**, on the other hand, are more specialized. Typical roles include:

- **Penetration Tester** - conducts authorized simulated attacks on systems to identify weaknesses.

- **Red Team Operator** - mimics advanced persistent threats to test an organization's full defensive readiness.

- **Vulnerability Assessor** - uses scanning tools to identify known vulnerabilities and misconfigurations.

- **Exploit Developer / Security Researcher** - discovers new vulnerabilities and develops proof-of-concept exploits, often contributing to the global cybersecurity community.

- **Bug Bounty Hunter** - works independently or through bounty platforms, finding vulnerabilities in exchange for rewards.

Ethical hackers are often employed by consulting firms, large tech corporations, financial institutions, or government agencies. Some work independently as contractors or freelance specialists.

Both paths share a strong job outlook. The cybersecurity workforce shortage is a global issue, with millions of unfilled positions projected over the next decade. Ethical hacking, while smaller in scale, remains in extremely high demand as organizations adopt "offensive security" as part of their strategy. In both cases, salaries are competitive, advancement opportunities are plentiful, and continuous learning ensures long-term career growth.

## Choosing the Right Path

Deciding between cybersecurity and ethical hacking ultimately depends on personal interest, aptitude, and long-term goals.

- **Mindset and Motivation:** Cybersecurity professionals tend to enjoy structure, risk management, and maintaining order, whereas ethical hackers thrive on curiosity, problem-solving, and creative challenge. If you like designing and enforcing defenses, cybersecurity may suit you better. If you prefer breaking and testing systems within legal limits, ethical hacking might be your passion.

- **Work Environment:** Cybersecurity roles often involve teamwork, documentation, and collaboration with multiple departments. Ethical hacking can be more independent and project-based, involving technical deep dives into specific systems.

- **Career Ambition:** If you aspire to leadership or policy roles, cybersecurity offers a broader platform. If you wish to be a hands-on technical expert, ethical hacking offers that niche specialization.

- **Learning Style:** Those who enjoy academic study and structured theory might prefer cybersecurity education, while those who enjoy hands-on experimentation may gravitate toward ethical hacking.

Both paths require adaptability, critical thinking, and integrity. The best professionals often blend both perspectives – understanding how attackers think (ethical hacking) while knowing how to defend effectively (cybersecurity).

## Conclusion

Cybersecurity and ethical hacking are two sides of the same coin in the mission to protect digital systems and information. Cybersecurity provides the **defensive shield**, while ethical hacking offers the **testing sword** that strengthens that shield through real-world challenges. Both careers are intellectually rewarding, socially valuable, and essential to modern society.

For students exploring this field, the key takeaway is that there is no wrong choice – only different directions within the same noble goal of securing cyberspace. A foundational understanding of computing, consistent practice, and a lifelong commitment to learning will ensure success in either path. Whether one becomes a **guardian** who builds defenses or a **tester** who challenges them, both roles contribute vitally to making the digital world safer for everyone.